\documentclass[12pt, a4paper]{article}
\usepackage[dvipdfmx]{graphicx}
\usepackage{amssymb}
\usepackage{amsmath}
\usepackage{bm}
\usepackage{color}
\usepackage{cite}
\usepackage{slashed}
\usepackage{subfigure}
\usepackage{epstopdf}            
\usepackage{epsfig}
            
\newcommand{\beq}{\begin{equation}}
\newcommand{\eeq}{\end{equation}}

\usepackage[colorlinks=true, linkcolor=black, citecolor=black,
urlcolor=black]{hyperref}

\begin{document}


\begin{titlepage}

\begin{flushright}
{IPMU18-0171}
\end{flushright}

\begin{center}

{\LARGE
{\bf 
Test of the $R(D^{(*)})$ anomaly at the LHC}
}

\vskip 2cm

Syuhei Iguro$^{1}$,
Yuji Omura$^{2}$
and
Michihisa Takeuchi$^{3}$

\vskip 0.5cm

{\it $^1$Department of Physics,
Nagoya University, Nagoya 464-8602, Japan}\\[3pt]

{\it $^2$
Kobayashi-Maskawa Institute for the Origin of Particles and the
Universe, \\ Nagoya University, Nagoya 464-8602, Japan}\\[3pt]

{\it $^3$ 
Kavli IPMU (WPI), UTIAS, University of Tokyo, Kashiwa, \\Chiba 277-8584, Japan}\\[3pt]

\vskip 1.5cm

\begin{abstract}
There are discrepancies between the experimental results and the Standard Model predictions,
in the lepton flavor universality of the semileptonic $B$ decays: $B \to D^{(*)} \ell \nu$.
As the new physics interpretations, new charged vector and charged scalar fields, that dominantly couple to the
second and third generations, have been widely discussed. In this paper, we study the
signals of the new particles at the LHC, and test the interpretations via the direct search for the new resonances.
In particular, we see that the $\tau \nu$ resonance search at the LHC has already covered most of the parameter regions favored by the Belle and BaBar experiments. We find that
the bound is already stronger than the one from the $B_c$ decay depending on the mass of charged scalar.
\end{abstract}

\end{center}
\end{titlepage}

\section{Introduction}

In 2012, the BaBar collaboration has reported that there are large discrepancies in the
lepton flavor universalities (LFUs) of the semileptonic $B$ decays: $B \to D \, \ell \, \nu$ and $B \to D^{*} \ell \, \nu$ ($\ell=e, \, \mu, \, \tau$).
The observables to measure the LFUs are defined as
\begin{equation}
R(D^{(*)}) = \textrm{Br}(B \to D^{(*)} \tau \, \nu)/\textrm{Br}(B \to D^{(*)} l \, \nu) ~(l=e, \, \mu),
\end{equation}
and the experimental results are $R(D)=0.440 \pm 0.072$ and $R(D^{*})=0.332 \pm 0.030$\cite{Lees2012xj,Lees2013uzd}. They largely deviate from
the Standard Model (SM) predictions: $R(D)_\textrm{SM}=0.299 \pm 0.003$ and $R(D^*)_\textrm{SM}=0.258 \pm 0.005$ \cite{Amhis2016xyh}.\footnote{See also Refs. \cite{Kamenik:2008tj,Fajfer2012vx,Lattice2015rga,Na2015kha,Bigi2016mdz,Bigi:2017jbd,Jaiswal:2017rve} .}
The $B$ decays associated with the light leptons are measured with good accuracy,
so that the branching ratios of $B \to D^{(*)} \tau \, \nu$ are larger than the SM predictions.
Interestingly, the Belle collaboration has also reported the excesses in $R(D^{(*)})$ \cite{Huschle:2015rga,Sato:2016svk,Hirose:2016wfn},
although the discrepancies are milder than the BaBar results.
Thus, it is expected that those excesses are the new physics signals and
there are new particles that couple to the SM fermions flavor-dependently. 
We note that the LHCb collaboration has also reported only $R(D^*)$ and the latest result is 
consistent with the SM prediction at the $1~\sigma$ level \cite{Aaij:2015yra,Aaij:2017uff}.
Recently, the Belle experiment also reported the data of the $D^*$ polarization in the $B \to D^{*} \tau \, \nu$ process, and the result is slightly deviated from the SM prediction at 1.5 $\sigma$ level \cite{Adamczyk}.

Motivated by the excesses, several new physics interpretations have been proposed.
One simple way to violate the LFU in the $B$ decay is to introduce a field that
couples to $\tau$ lepton. The field needs to couple to the heavy quarks, bottom and
charm quarks, as well.
One good candidate for such a field is a charged scalar, $H_{\pm}$, that has Yukawa couplings with the heavy quarks and heavy leptons \cite{Crivellin2012ye,Celis:2012dk,Tanaka2012nw,Ko:2012sv,Crivellin:2013wna,Crivellin:2015hha,Kim:2015zla,Cline:2015lqp,Ko:2017lzd,Iguro:2017ysu,Fuyuto:2017sys,Iguro:2018qzf,Iguro:2018oou,Martinez:2018ynq,Fraser:2018aqj,Li:2018rax}. The Yukawa couplings are, in general, flavor-dependent, so that
we can assume that the couplings with bottom, charm and $\tau$ lepton are relatively large, compared to the other elements. Then, the exchange of a charged scalar at the tree-level induces the violation of the LFU.
This simple scenario has been proposed just after the announcement of the BaBar result,
and the way to prove the new physics directly/indirectly has been widely discussed. 

Instead of the charged scalar, we can discuss a charged vector, $W^\prime_{\pm}$, that dominantly couples to the second
and third generations \cite{He:2012zp,Greljo:2015mma,Boucenna:2016wpr,He:2017bft,Cvetic:2017gkt,Asadi:2018wea,Greljo:2018ogz}. In order to introduce such a vector field, additional gauge symmetry is required
and the SM gauge symmetry may be extended. In addition, a non-trivial setup would be necessary 
to make the $W^\prime_{\pm}$ couplings flavor-dependent. For instance, we can discuss a gauged flavor symmetry
or we can expect that some heavy fermions effectively induce the flavorful couplings according to the mass mixing
with the SM fermions.

In this paper, we focus on those two new physics interpretations and discuss
the consistency with the direct search for the new phenomena at the LHC in the each setup.
In particular, it is recently claimed that the charged scalar explanation is in tension with the $B_c$ decay \cite{Alonso:2016oyd,Akeroyd:2017mhr,Li:2016vvp}. 
We study the $\tau \nu$ resonance search at the LHC and see that
the bound is stronger than the one from the $B_c$ decay. 

We summarize the each explanation in Sec. \ref{explanation}:
the $H_\pm$ case in Sec. \ref{H} and the $W^\prime_\pm$ case in Sec. \ref{Wprime}.
Based on the results in Sec. \ref{explanation}, we study the signals
at the LHC in Sec. \ref{LHC}. 
Sec. \ref{LHC-H} and Sec. \ref{LHC-Wprime} devote to the analyses of
the  $H_\pm$ and $W^\prime_\pm$ scenarios, respectively.
We summarize our results in Sec. \ref{summary}.

\section{The explanation of the $R(D^{(*)})$ anomaly}
\label{explanation}

There are large discrepancies between the experimental results
and the SM predictions in the LFUs of the semileptonic $B$ decays: $B \to D^{(*)} \ell \nu$.
In the SM, the processes are given by the tree-level diagrams. Then,
relatively large new interaction is required to compensate the SM contribution. 
If there is a heavy charged particle that couples to quarks and leptons 
flavor-dependently, the following operators could be generated by the heavy particle exchange:
\begin{eqnarray}
{\cal H}_{eff}&=&(C^V_{SM}+C^V_{L}) (\overline{b_L} \gamma_\mu c_L) (\overline{\nu_{\tau L}} \gamma^\mu \tau_L)+C^V_{R} (\overline{b_R} \gamma_\mu c_R) (\overline{\nu_{\tau R}} \gamma^\mu \tau_R) \nonumber  \\
&&+C^S_L (\overline{b_R}  c_L) (\overline{\nu_{\tau L}} \tau_R)+C^S_R (\overline{b_L}  c_R) (\overline{\nu_{\tau L}} \tau_R) +h.c.,
\label{Heff}
\end{eqnarray}
here $C^V_{SM}$ expresses a SM contribution generated by $W$ boson, with $C^V_{SM}=4G_FV_{cb}^*/\sqrt2$. 
The two terms in the fist line can be generated by the $W^\prime$ exchange.\footnote{The left-right mixing operator, $(\overline{b_R} \gamma_\mu c_R) (\overline{\nu_{\tau L}} \gamma^\mu \tau_L)$, would be allowed in general, but we assume that it is not generated by our $W^\prime$ since the $W$-$W^\prime$ mixing is suppressed. }
The last two terms can be from the $H_\pm$ exchange.
In this paper, we focus on these two scenarios with the SU(3)$_c$-singlet mediators.
In the following subsections, we review the each new physics scenario and estimate the size of coefficient required by the excesses.

\subsection{Charged scalar case}
\label{H}
To begin with, we discuss a possibility that charged scalar, $H_\pm$, resides behind the $R(D^{(*)})$ anomalies.
The charged scalar can be introduced by adding extra Higgs SU(2)$_L$ doublets.
The Yukawa couplings between $H_\pm$ and the SM fermions depend on the setup, but in general 
the scalar couples to all of the SM fermions. Most of the Yukawa couplings are strongly constrained by the
flavor physics, so that we have to assume a specific alignment of the couplings. 
In fact, we limit our Yukawa couplings to those between 2nd and 3rd generations as in \cite{Iguro:2017ysu}.
Assuming such a specific
parameter choice, we can focus on the $b \to c $ transition induced by the Yukawa coupling of charged scalar, i.e.,
\beq
{\cal L}_{H_\pm}= - H_- \left \{ Y_R  (\overline{b_R} c_L) + Y_L  (\overline{b_L} c_R) + Y_\tau(\overline{\tau_R}  \nu_{\tau L}) \right \} +h.c..
\eeq
As mentioned above, $H_\pm$ is originated from $SU(2)_L$-doublet scalars, and the neutral components of 
the doublet also appear as physical fields, after the electroweak symmetry breaking.
The masses of the neutral scalars are expected to be around the charged scalar mass ($M_H$) to evade the constraint from the electroweak precision test \cite{Lee:1973iz}. 
The mixing between a SM Higgs and the additional neutral scalar in this scenario should be tiny \cite{Iguro:2018qzf}.  
Then the Yukawa couplings involving the neutral ones are also evaluated as $Y_L$ and $Y_R$, approximately. 
When one neutral scalar is denoted as $H_0$, the couplings between $H_0$ and
the down-type (up-type) quarks are described as $Y_R V $ $(V Y_L ) $, where $V$ is the CKM matrix.

Integrating out $H_\pm$, we obtain 
\begin{equation}
\label{H-coupling}
{\cal H}_{H_\pm}=-\frac{Y_L Y^*_\tau}{M_{H}^2} (\overline{b_L}  c_R) (\overline{\nu_{\tau L}} \tau_R)-\frac{Y_R Y^*_\tau}{M_{H}^2} (\overline{b_R}  c_L) (\overline{\nu_{\tau L}} \tau_R),
\end{equation} 
where $M_H$ is the charged scalar mass.
Then, we find that $Y_R $ is also strongly constrained by the $B_s$-$\overline{B_s}$ mixing, taking into account the 
neutral Higgs exchange at the tree level.
As a result $Y_R $ is not useful to improve $R(D^{(*)})$ \cite{Iguro:2017ysu}. 
We assume $|Y_R| \ll |Y_L|$ in our analysis below.
If the charged scalar mass is less than a few TeV, these operators largely contribute to the semileptonic $B$ decay.

The numerical descriptions of $R(D)$ and $R(D^*)$ are given by \cite{Tanaka2012nw}

\begin{align}
R(D)\simeq R(D)_{SM}\biggl{\{}1+1.5{\rm Re}\bigl[ C^{\prime S}_L+C^{\prime S}_R\bigl]+|C^{\prime S}_L+C^{\prime S}_R|^2\biggl{\}},
\end{align}
\begin{align}
R(D^*)\simeq R(D^*)_{SM}\biggl{\{}1-0.12{\rm Re}\bigl[ C^{\prime S}_L-C^{\prime S}_R\bigl]+0.05|C^{\prime S}_L-C^{\prime S}_R|^2\biggl{\}},
\label{HpRDS}
\end{align}
where $C^{\prime S}_{I}$  ($I=L$, $R$) is a normalized coefficient given as $C^{\prime S}_{I}=C^{S}_{I}/C^V_{SM}$, with $C_L^S=-Y_LY_\tau^*/m_H^2$, $C_R^S=-Y_RY_\tau^*/m_H^2$. 
The required value to achieve the excess of the world average within $1~\sigma$ is estimated as
\beq
\label{charged}
\frac{\left | Y_L Y^*_\tau \right |}{M_{H}^2} \simeq 1.38\times10^{-6}~{[{\rm GeV}^{-2}]},
\eeq
where, we select the phase of $Y_L Y^*_\tau$ to minimize the $\chi^2$.

The explanation of $R(D^*)$ is, however, constrained indirectly by the $B_c$ decay \cite{Alonso:2016oyd,Akeroyd:2017mhr}.
The $B_c$ meson decay is easily enhanced by the scalar-type operator. The leptonic decay, $B_c \to \tau \nu$,
is still not observed, but the total decay width and the hadronic decay have been measured and the results
are consistent with the SM predictions, although the observables suffer from the large theoretical uncertainty. 
The authors of Refs. \cite{Alonso:2016oyd,Akeroyd:2017mhr} derive the upper bound on the leptonic decay, 
taking into account the uncertainty, and obtain the upper bound $R(D^*) \lesssim 0.27$ in the charged scalar scenario.
The LHCb collaboration has recently reported the consistent $R(D^*)$ results with the SM prediction. 
The enhancement of $R(D)$ while keeping the $R(D^*)$ consistent to the SM prediction
would be only achieved by tuning the phase of $Y_L Y^*_\tau$~\cite{Iguro:2017ysu}.

\subsection{$W^\prime$ case}
\label{Wprime}
We can discuss the possibility that the coefficients in Eq. (\ref{Heff}) are induced by
the heavy vector boson exchange. 
In the extended SM with extra non-abelian gauge symmetry, massive extra gauge bosons are predicted.
If the SM quarks are charged under the extra gauge symmetry, an extra charged gauge boson, $W^\prime$,
may couple to the third-generation quark and lepton as
\beq
\label{Wprime-coupling}
{\cal L}_{W^\prime_I}= W^\prime_{I\mu} \left \{ g_I  (\overline{b_I} \gamma^\mu c_I)  + g_{I \, \tau}(\overline{\tau_I} \gamma^\mu \nu_{\tau I}) \right \}+h.c.,
\eeq
where, $I$ denotes the chirality: $I=L, \, R$.
The couplings $g_I$ and $g_{I \, \tau}$ depend on the detail of the setup, and the other couplings involving light SM fermions
may arise at the low energy.
Assuming the third-generation couplings are dominant, 
we expect the following operators induced:
\begin{align}
{\cal H}_{W^\prime}=&\frac{g_L g^*_{L \, \tau}}{M^2_{W^\prime_L}} (\overline{b_L} \gamma_\mu c_L) (\overline{\nu_{\tau L}} \gamma^\mu \tau_L)+\frac{g_R g^*_{R \, \tau}}{M_{W^\prime_R}^2} (\overline{b_R} \gamma_\mu c_R) (\overline{\nu_{\tau R}} \gamma^\mu \tau_R)
\end{align} 
where $M_{W^\prime_I}$ denotes the $W^\prime_I$ mass.
$R(D)$ and $R(D^*)$ are numerically evaluated as \cite{Asadi:2018wea}
\begin{align}
R(D^{(*)})\simeq R(D^{(*)})_{SM}\biggl{\{}|1+C_{L}^{\prime V}|^2+|C_{R}^{\prime V}|^2\biggl{\}},
\end{align}
where $C^{\prime V}_{I}$ ($I=L$, $R$) is a normalized coefficient given as $C^{\prime V}_{I}=C^{V}_{I}/C^V_{SM}$ with $C_L^V=g_Lg_{L\tau}^*/m_{W_L^\prime}^2$, $C_R^V=g_Rg_{R\tau}^*/m_{W_R^\prime}^2$. 
In order to accommodate the $R(D^{(*)})$ excesses within $1~\sigma$ of the world average, 
the required value of the each coefficient is estimated as 
\beq
\label{eq:grelation2}
\frac{g_L g^*_{L \, \tau}}{M^2_{W^\prime_L}} \simeq 1.07\times 10^{-7}~ {\rm{[GeV^{-2}]}},~\frac{g_R g^*_{R \, \tau}}{M^2_{W^\prime_R}} \simeq 5.55\times 10^{-7}~ {\rm{[GeV^{-2}]}}.
\eeq

\section{Test of the new physics at the LHC}
\label{LHC}
In this section, we study the signal of the each scenario at the LHC
based on the  above discussion.
In our models, the charged resonances ($V^+ = W^{\prime +}_I, H_+$) are produced in association with the third-generation quark
and decay to $\tau \nu$ and $bc$ as follows\footnote{We work in 4 flavor scheme.}: 
\begin{eqnarray}
g \, c &\to&  V^+ \, b \to  \tau^+ \nu \, b.
\end{eqnarray}

The searches for heavy $\tau \nu$ resonances have been performed at the LHC both in ATLAS and CMS experiments~\cite{Khachatryan:2015pua,CMS:2016ppa,Aaboud:2018vgh,Sirunyan:2018lbg}, and severely constrain various models.\footnote{Our model focus on the heavy resonance decaying into $\tau \nu$ and it is 
different from the search for leptoquark using $b + \tau \nu$ in the final state discussed in Ref.~\cite{Altmannshofer:2017poe}. See also Refs.~\cite{Iguro:2017ysu,Abdullah:2018ets}.}
We numerically found the analysis reported by the CMS collaboration using the data at the LHC Run II with 35.9 fb$^{-1}$~\cite{Sirunyan:2018lbg}
sets the most stringent bound on our models, where they focus on the $W^\prime$ heavier than 400 GeV with 
the universal couplings to quarks of all generations. 

Since the heavy resonances  are only couples to the third generation in our models and 
the spin structures of the $H_\pm$ and the $W^\prime$ are different, the efficiency and the acceptance 
of the selection cut should be estimated by the simulation.
We calculate the acceptances for the case of  $H_\pm$ and $W^\prime$
using {\tt MadGraph5}~\cite{Alwall:2014hca} and {\tt PYTHIA8}~\cite{Sjostrand:2006za}.
The generated events are interfaced to {\tt DELPHES3}~\cite{deFavereau:2013fsa} 
for the fast detector simulation. 

We follow the event selection cuts exploited in~\cite{Sirunyan:2018lbg} as follows:
\begin{itemize}
\item exactly one $\tau$-tagged jet, satisfying $p_{T,\tau}\ge$~80GeV and $|\eta_\tau|\le2.4$, 
\item no isolated electrons nor muons ($p_{T,e},p_{T,\mu} \ge$ 20GeV, $|\eta_e| \le2.5$, $|\eta_\mu|\le2.4$),
\item large missing momentum $E\!\!\!/_T\ge 200$~GeV, 
\item and it is balanced to the $\tau$-tagged jet: 
$\Delta \phi(E\!\!\!/_T, \tau)\ge$2.4 and 
$0.7\le p_{T,\tau}/E\!\!\!/_T \le1.3$, \\
where $\Delta \phi(E\!\!\!/_T, \tau)$ is the azimuthal angle between the missing momentum and the $\tau$-jet.
\end{itemize}

\begin{figure}[h]
  \begin{center}
    \includegraphics[width=7cm]{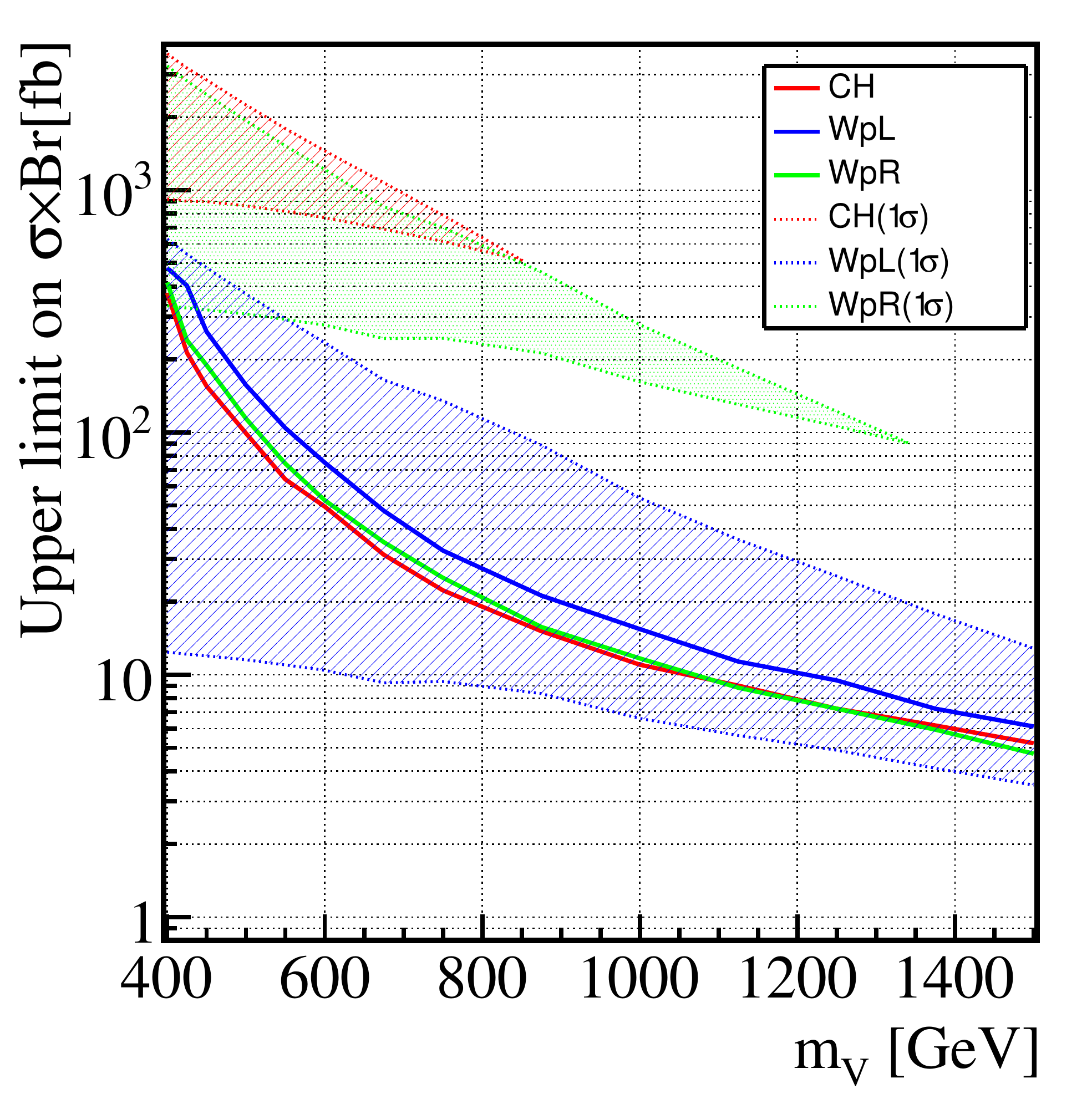}
    \caption{The upper bound on the cross section times branching ratio in the each model at 95$\%$ CL. 
See the main text for the description.
    }
    \label{fig:LHCconstraints} 
  \end{center}
\end{figure}

After the above event selection cuts, we plot the $m_T$ distribution and performed the binned log-likelihood analysis using the background $m_T$ distribution\cite{Sirunyan:2018lbg}.   
We show the resulting 95~\% CL upper bound on the signal cross section times its branching ratio to $\tau\nu$ mode as a function of the resonance mass $m_V$ for each model in Figure~\ref{fig:LHCconstraints}. 
The difference of the spin structure provides different upper bounds, for charged scalar (red thick solid), $W_L^\prime$ (blue thick solid), and
 $W_R^\prime$ (green thick solid).
The constraint on the charged scalar case is more stringent than the other cases in most of the $m_V$ region because of the harder $m_T$ distribution.

We also overlay the expected signal cross sections in the same plot 
for the three cases, in the red hatched region ($H_\pm$), in the blue hatched region ($W^\prime_L$) and in the green hatched region ($W^\prime_R$),
assuming the couplings are compatible to accommodate the $R(D^{(*)})$ observation in 1$\sigma$ level.
In our models, we can parametrize the cross section with $(M, g, g_\tau)$, where $g$ is the $c$-$b$-$V$ 
coupling and can be taken real without loss of generality; e.g., $(M, g, g_\tau)=(M_H, Y_L, Y_\tau)$, 
and $(M, g, g_\tau)=(M_{W^{\prime }_I}, g_I, g_{I \, \tau})$, respectively. 
The cross section for the above process is given as follows:
\beq
\label{Htotaunu}
\sigma (pp \to V^\pm ) \times Br(V^\pm \to \tau \nu) 
= \sigma_0(m_V) \times \frac{|g|^2 |g_\tau|^2}{3|g|^2 +|g_\tau|^2} = \sigma_0(m_V) \times \bar{g}^2 \frac{r}{3 + r^2}. 
\eeq 
Here, we define the variables, $\bar{g}^2=|g \, g_\tau|$,
which is related to the $R(D^{(*)})$ prediction, 
and $r=|g_\tau/g|$.
Changing $r$, while keeping $\bar{g}$, varies the cross section.
We find that basically, the signal cross section is maximized at $r=\sqrt{3}$ when we fix $\bar{g}$.
We also impose the perturbativity of the couplings, i.e. $|g|,|g_{\tau}| \leq 1$, and $r$ would be constrained as well. 
We assume $\bar{g}^2= a_V \, M_V^2$  to accommodate the $R(D^{(*)})$ observation at 1 $\sigma$, 
that is $a_{H_\pm}=1.36\times 10^{-6} {\rm GeV}^{-2}$, and $a_{W^\prime_L}=1.07\times 10^{-7} {\rm GeV}^{-2}$.
Compared with the required $a_{W^\prime_L}$,  $a_{H_\pm}$ needs to be large to accommodate the excess. 
This stems from small coefficients in Eq. (\ref{HpRDS}). 
Then we expect the more stringent bound can be obtained for a charged scalar scenario. 
Since we impose $\bar{g}^2 \le 1$ due to the perturbativity, $M_V$ exhibits an upper bound 850~GeV for $H^+$ and 3~TeV for $W^\prime_L$.
The upper boundary of the hatched region is given by $r=\sqrt{3}$ for $\bar{g}^2 \le 1/\sqrt{3}$ or $r=1/\bar{g}^2$ ($g_\tau$=1) for $\bar{g}^2 > 1/\sqrt{3}$. 
The lower boundary of the region is given by $r=\bar{g}^2$ ($|g|=1$). Note that the lower boundary depends on the upper bound of the allowed couplings.

In the following, we discuss the current status in more detail and the compatibility with the $R(D^{(*)})$ enhancements for each case.

\subsection{Charged scalar case}
\label{LHC-H}
First, we investigate the charged scalar signal at the LHC.
As we discuss in Sec. \ref{H}, the $R(D^{(*)})$ anomaly requires
the sizable interaction between the heavy quarks and the $\tau$ lepton.
In our study of the charged scalar case, there are four free parameters:
\beq
M_H,~Y_L,~ {\rm Re}\left ( Y_\tau \right ),~{\rm Im}\left ( Y_\tau \right ).
\eeq
The explanation of the $R(D^{(*)})$ anomaly fixes one combination of them, 
as shown in Eq.~(\ref{charged}).
Tuning the phase of $Y_\tau$, the bound from the $B_c$ decay
can be evaded, although the enhancement of $R(D^*)$ is suppressed.
Note that the magnitudes of the Yukawa couplings
are not so large and the total decay width is small enough in our study.
The production cross section of $H_\pm$ only depends on $Y_L$,
while $|Y_\tau|$ contributes to the branching ratio of $H_\pm$ only.
The production cross section times the branching ratio is
given by substituting $(M, g, g_\tau)=(M_H, Y_L, Y_\tau)$ in Eq. (\ref{Htotaunu}).
The exclusion line and the predicted region of the charged scalar case are
shown in Fig. \ref{fig:LHCconstraints}. As we see the red line and the red hatched region in Fig. \ref{fig:LHCconstraints}, the region to accommodate the experimental results on $R(D^{(*)})$ within the $1 \sigma$ level is already excluded by this $\tau \nu$ resonance search. The light charged scalar region, e.g. $M_H \lesssim 400$ GeV, could achive the experimental results within the $2 \sigma$ level.

\begin{figure}[h]
  \begin{center}
    \includegraphics[width=7.85cm]{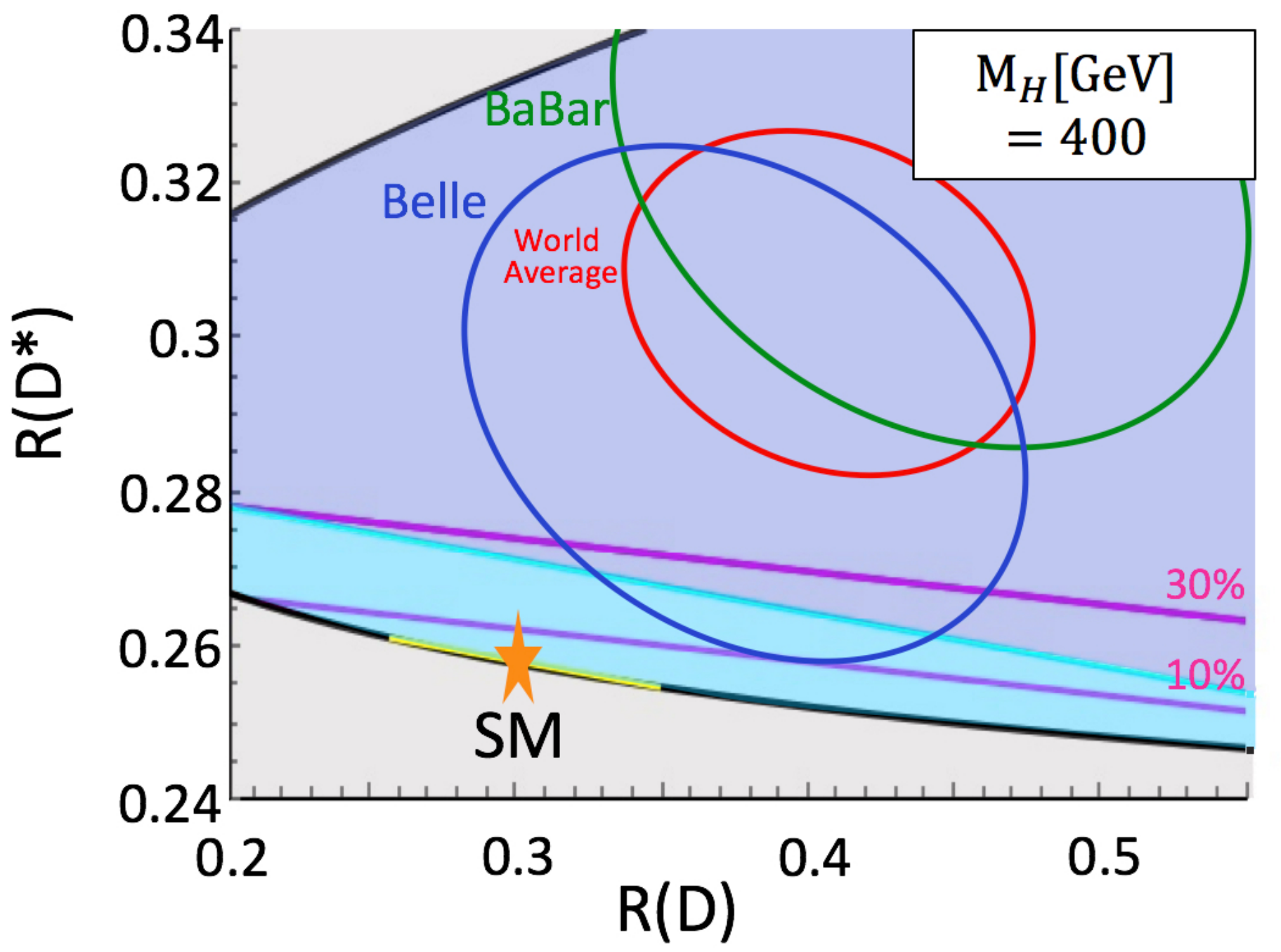}
    \includegraphics[width=7.85cm]{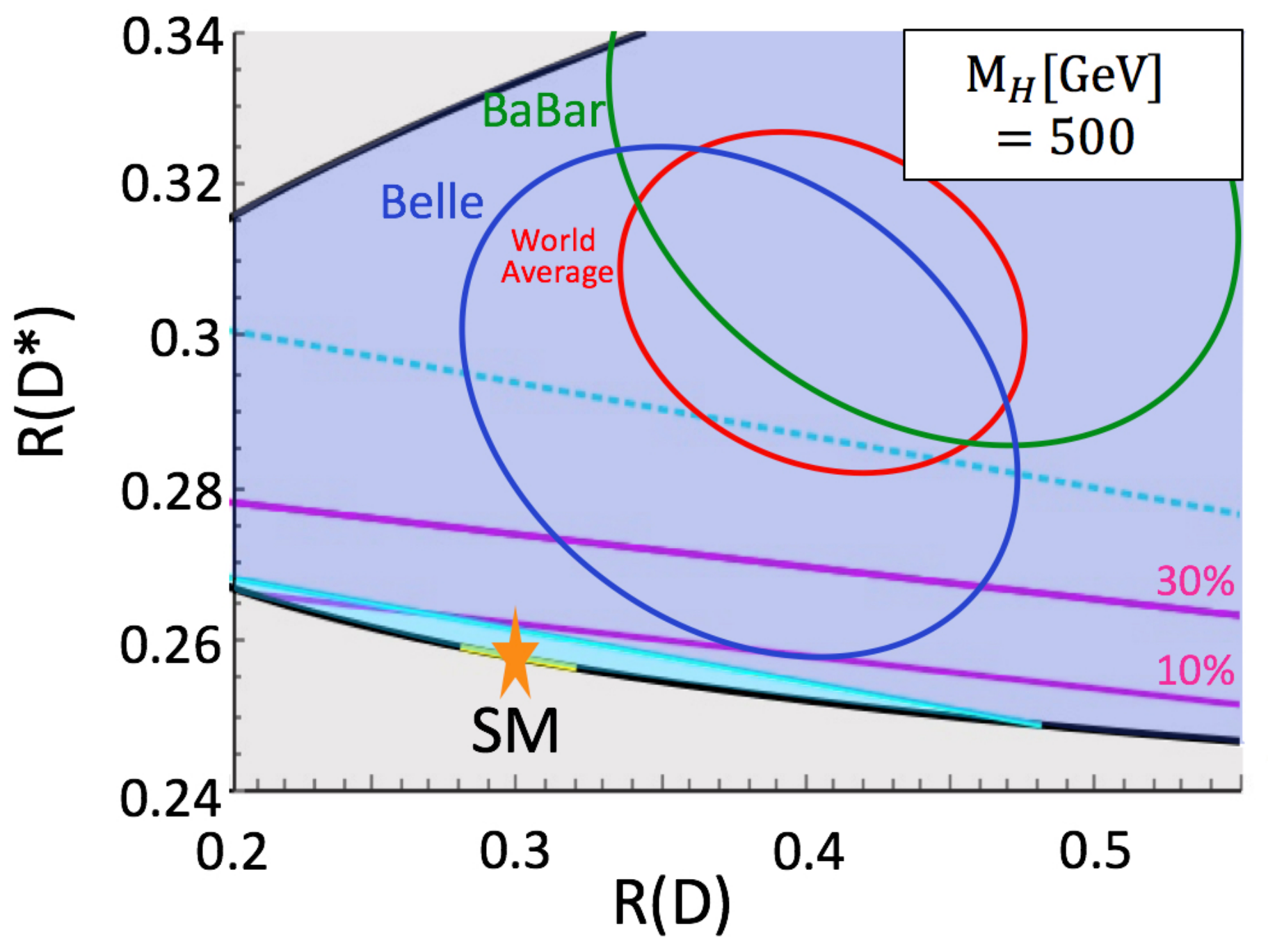}
    \includegraphics[width=7.85cm]{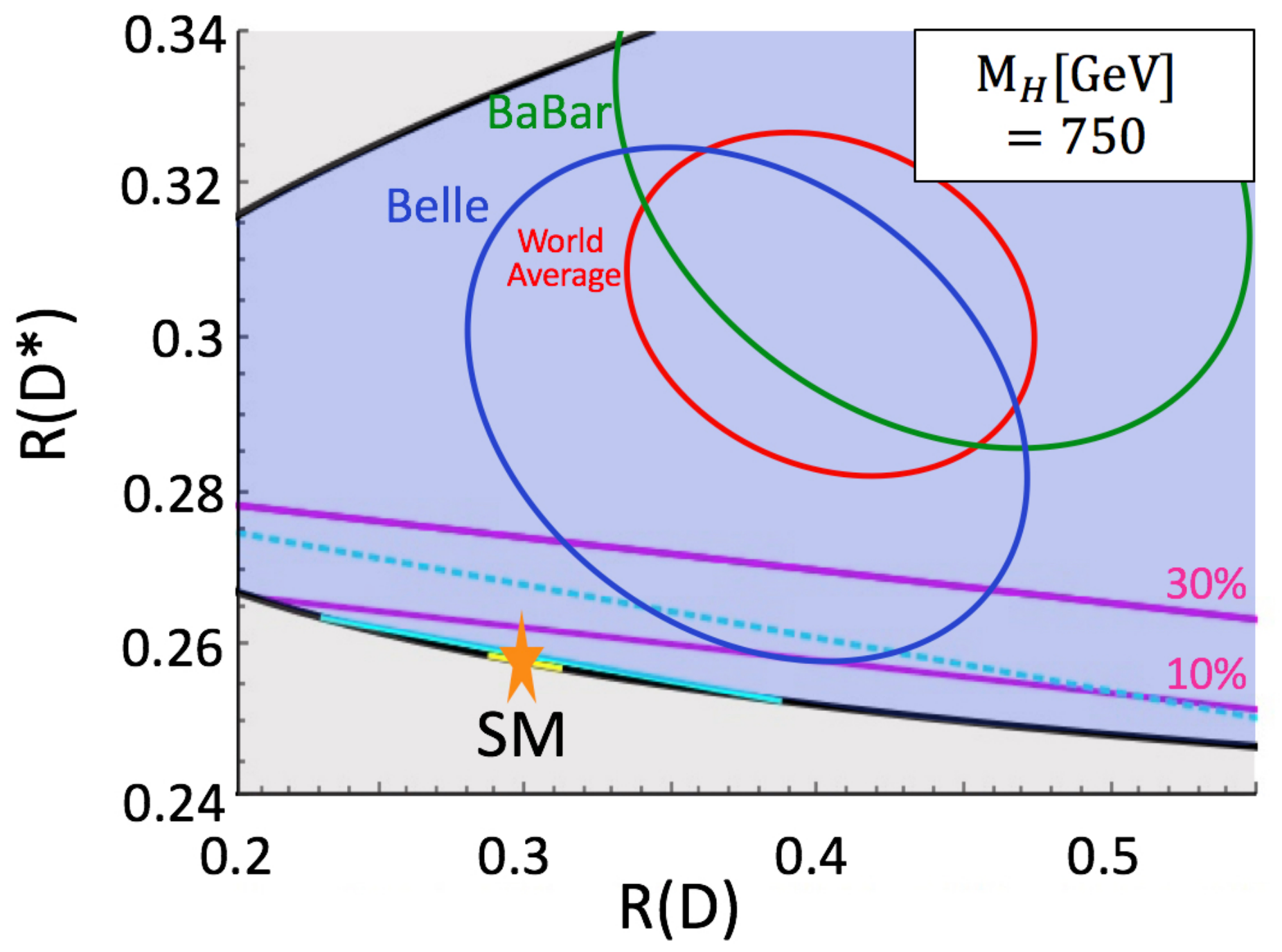}
    \includegraphics[width=7.85cm]{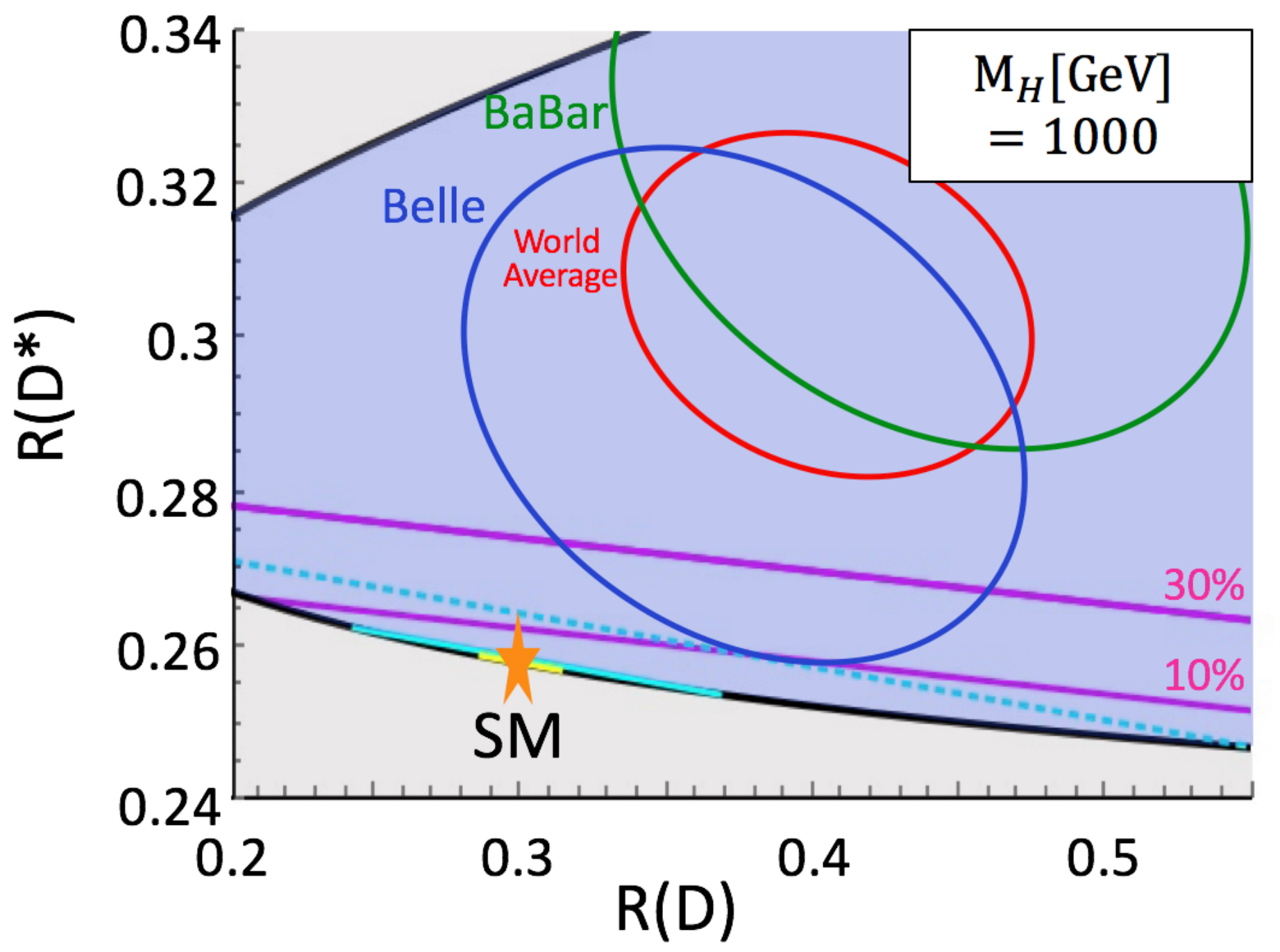}
    \caption{$R(D)$ vs. $R(D^*)$ with the constraints from the $B_c$ decay (magenta lines) and from the charged heavy resonance search at the LHC. The charged scalar masses are fixed at 400 GeV, 500 GeV, 750 GeV and 1000 GeV, as denoted on the each panel. The gray region is out of our prediction and the region inside of the black lines is realized by taking appropriate $Y_L$ and $Y_\tau$.
Note that the blue, green and red ellipses correspond to the 1 $\sigma$ regions reported by the Belle, BaBar and the HFLAV collaborations.
The SM prediction for $R(D)$ and $R(D^*)$ is marked with the asterisk. 
The cyan dashed lines correspond to the predictions by taking $|Y_L|=\pi$.
In the light purple region, our model is totally excluded by the $\tau \nu$ resonance search. In the light blue region, our model can be allowed by tuning $Y_L$ and $Y_\tau$.  }
    \label{Figure-H}
  \end{center}
\end{figure}

Fig.~\ref{Figure-H} shows our predictions on the $R(D)$ and $R(D^*)$ plane for the fixed charged scalar mass, 400 GeV, 500 GeV, 750 GeV and 1000 GeV. The gray region is out of our prediction and the region inside of the black lines is realized by taking $Y_L$ and $Y_\tau$ appropriately.
 Note that the blue, green and red ellipses correspond to the 1 $\sigma$ regions reported by the Belle, BaBar and the HFLAV collaborations.
The SM prediction for $R(D)$ and $R(D^*)$ is also marked with the asterisk. 
In Fig. \ref{Figure-H}, we also draw the constraints from the $B_c$ decay by the thick magenta lines.
Above the magenta line, the leptonic decay of $B_c$ is larger than 30 \% and 10 \%,
so that the region above the line is excluded indirectly. 
The cyan dashed lines correspond to the predictions by taking $|Y_L|=\pi$.\footnote{We draw them for an illustration although such a relatively large coupling does not respect the narrow width assumption.}

The bound from the heavy resonance search at the LHC is our main topic in this paper.
We study the excluded parameter region on the $R(D)$ and $R(D^*)$ plane based on Fig.~\ref{fig:LHCconstraints}.
As the signal cross section depends both on $Y_L$ and $Y_\tau$, we have one additional degree of freedom than the product $|Y_L Y_\tau|$.
The charged higgs scenario is totally excluded by the $\tau \nu$ resonance search in the light purple region while 
it would be allowed by tuning the ratio $Y_\tau/Y_L$ in the light blue region.
Interestingly, we find that most of parameter region is already excluded by the LHC searches.
Only for the light charged scalar, $m_{H} \lesssim 450$ GeV, 
we can find a parameter region consistent at 1 $\sigma$ by allowing a large $|Y_L|=\pi$.
This is because the LHC $\tau \nu$ resonance constraints set more stringent constraints for the heavier resonance.
We also stress that the bound by the search at the LHC is stronger
than the one by the $B_c$ decay for the considered range $m_H \ge 400$~GeV.
 In Fig. \ref{fig:coupling}, we also study the excluded parameter region on the $Y_L$ and $Y_\tau$ plane based on Fig.~\ref{fig:LHCconstraints}. 
 The figure on the upper left is for the $H_\pm$ scenario. 
 In the figure, the red line is for the mass of 400GeV.
 The dotted line express the combination of couplings that accomodate the 1 $\sigma$ of $R(D^{(*)})$. The shaded region is excluded by the collider bound.
    As is shown in Fig. \ref{Figure-H}, the $H_\pm$ scenario is constrained stringently.  The parameter region for the $H_\pm$ scenario is completely excluded.

\subsection{$W^\prime$ case}
\label{LHC-Wprime}
Next, we study the $W^\prime$ scenario.
In the Eq. (\ref{Wprime-coupling}), there are two types of the gauge couplings with 
the fermions: one is the coupling with right-handed fermions and the other is the one with left-handed fermions.
It depends on the charge assignment of the extra non-abelian gauge symmetry.
If the extra SU(2) symmetry is assigned to the left-handed fields like the
SM SU(2)$_{L}$, $W_L^\prime$ only couples to the left-handed fields as well.
In a model with $W^\prime_R$, SU(2)$_R$ symmetry would be assigned to the right-handed fermions. 
In those cases, $W^\prime_I$ originated from SU(2)$_{I}$ may mix with $W$ from SU(2)$_{L,SM}$. 
The mixing is strongly constrained by the electroweak precision observables, so the mixing should be very tiny. 
We mainly discuss the case with $W_L^\prime$ in the following for the demonstration. The one with $W_R^\prime$ can be treated in the same way.  

In the same way to the charged scalar case, we discuss the $\tau \nu$ resonance originated from the on-shell $W_L^\prime$.
Our relevant parameters are
\beq
M_{W^\prime},~g_L,~ {\rm Re}\left ( g_{L \, \tau} \right ),~{\rm Im}\left ( g_{L \, \tau} \right ).
\eeq

In the same way as the $H_\pm$ case,
the cross section is given by substituting $(M, \, g, \, g_\tau)=(M_{W^\prime}, \, g_L, \, g_{L \, \tau})$ in Eq. (\ref{Htotaunu}).
The exclusion line and the predicted region of the $W_L^\prime$ case are
shown in Fig. \ref{fig:LHCconstraints}. As we see the blue line and the blue hatched region in Fig. \ref{fig:LHCconstraints}, the constraint is not so tight. 
The exclusion line and the prediction of the $W_R^\prime$ case are also shown as the green line and the green hatched region. 
As far as the gauge coupling is less than one, the $W_R^\prime$ case is almost excluded unless $m_{W^\prime_R}$ is lighter than 420~GeV,
since the gauge coupling coupling required to accommodate $R(D^{(*)})$ anomaly is relatively large
compared to the one in the $W_L^\prime$ case.
Note that there is no bound from the $B_c$ decay, since the operator is given by the vector-vector coupling in this scenario.

\begin{figure}[h]
  \begin{center}
    \includegraphics[width=7.5cm]{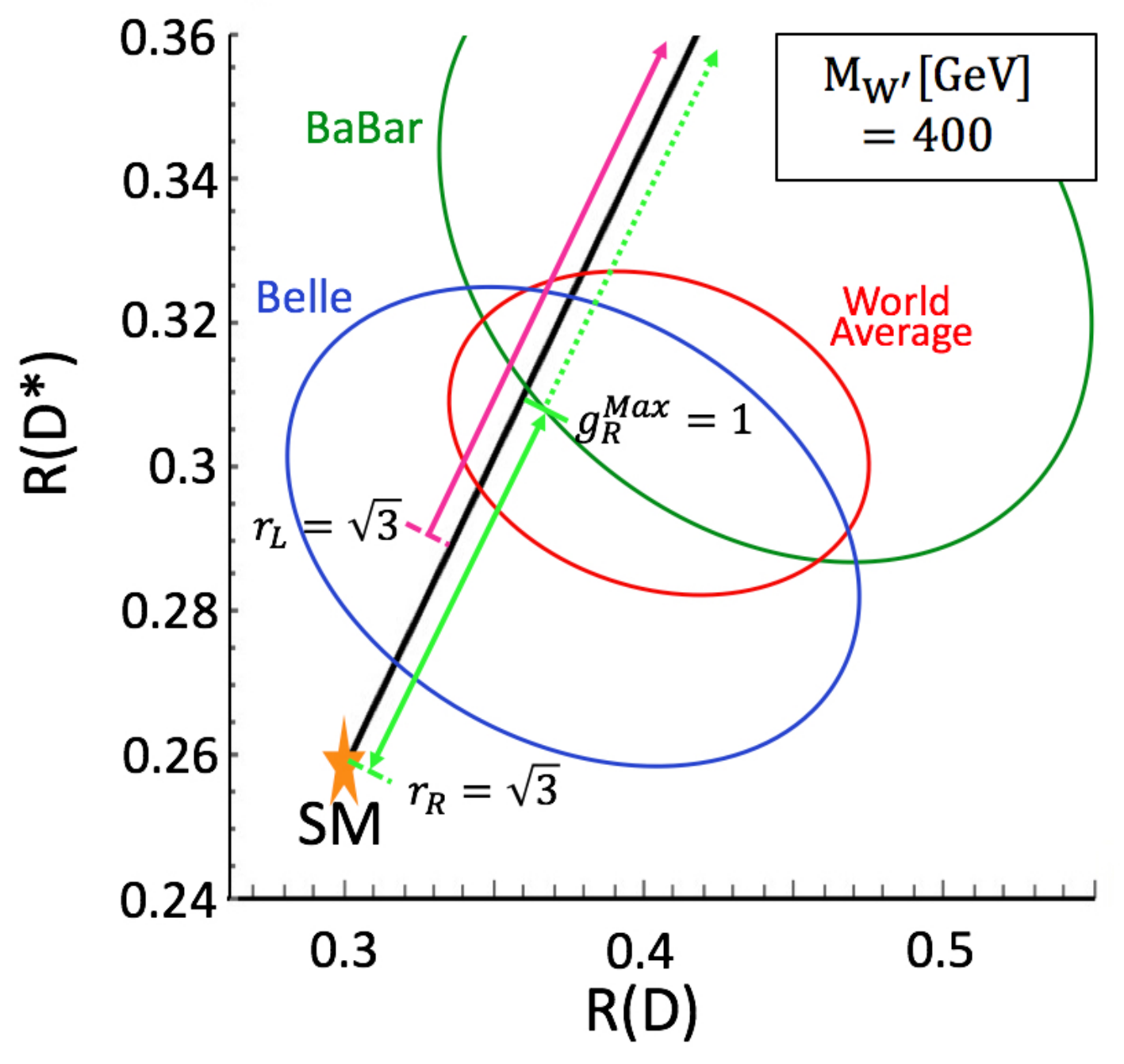}
    \includegraphics[width=7.5cm]{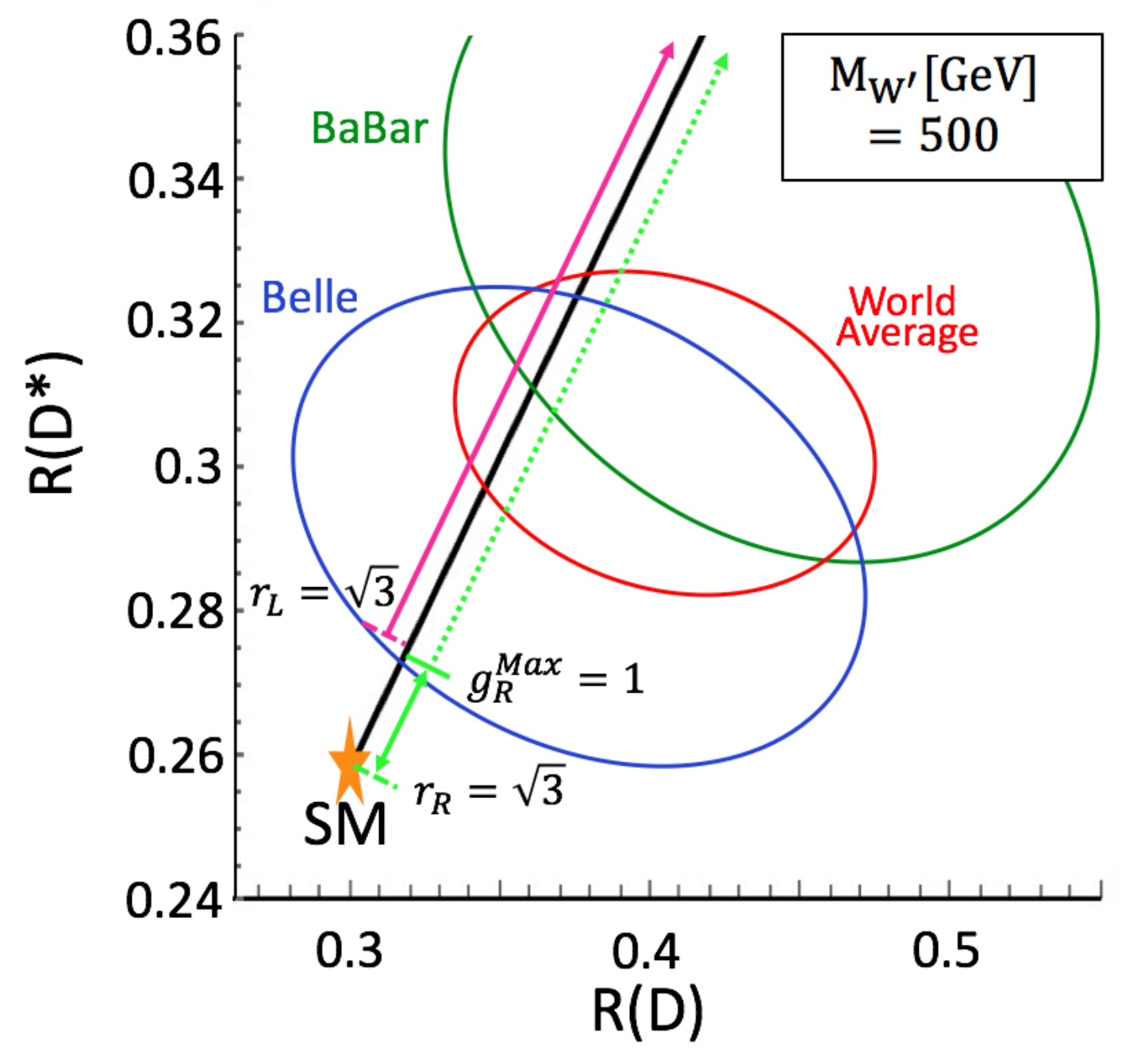}
    \includegraphics[width=7.5cm]{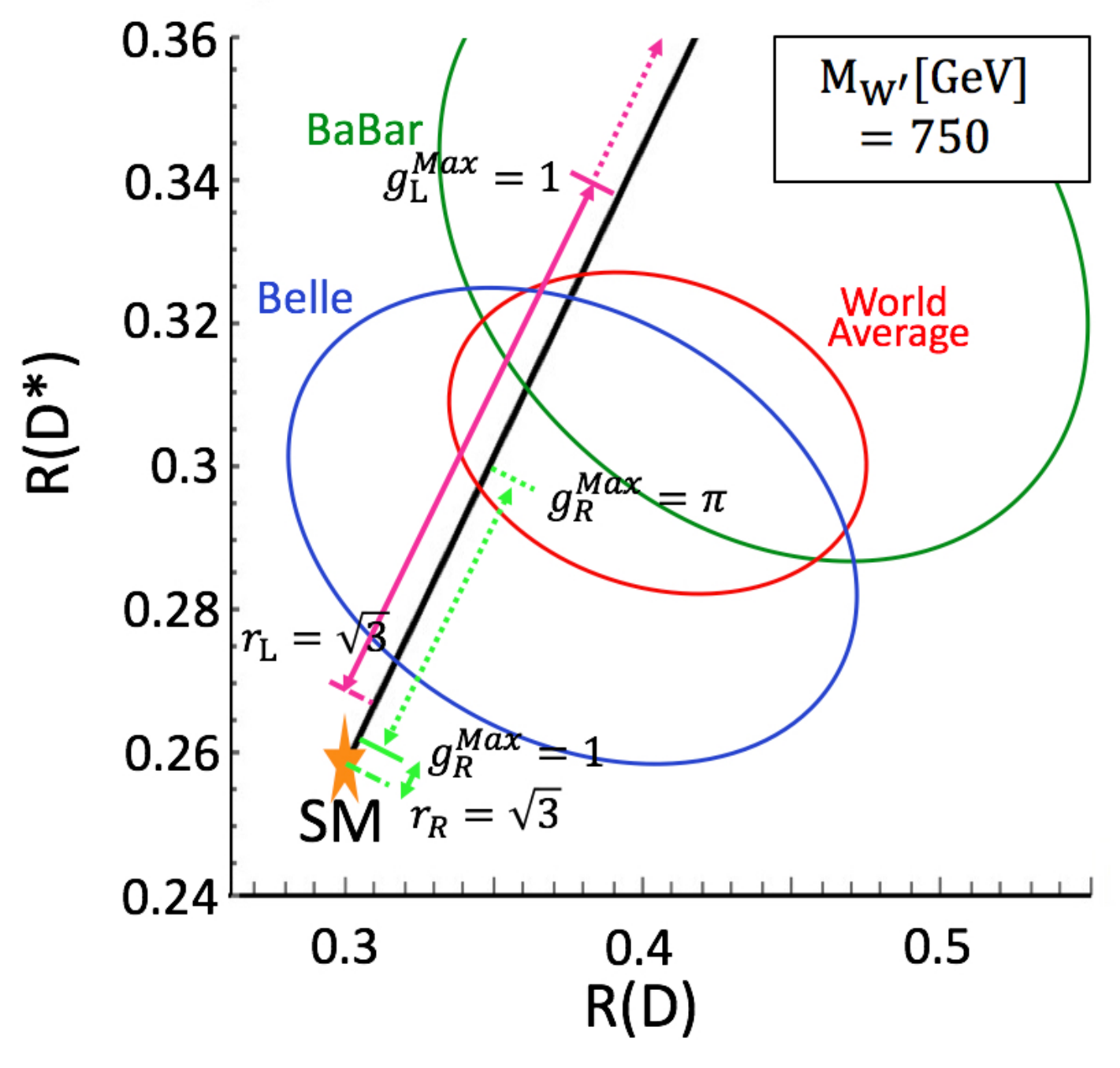}
    \includegraphics[width=7.5cm]{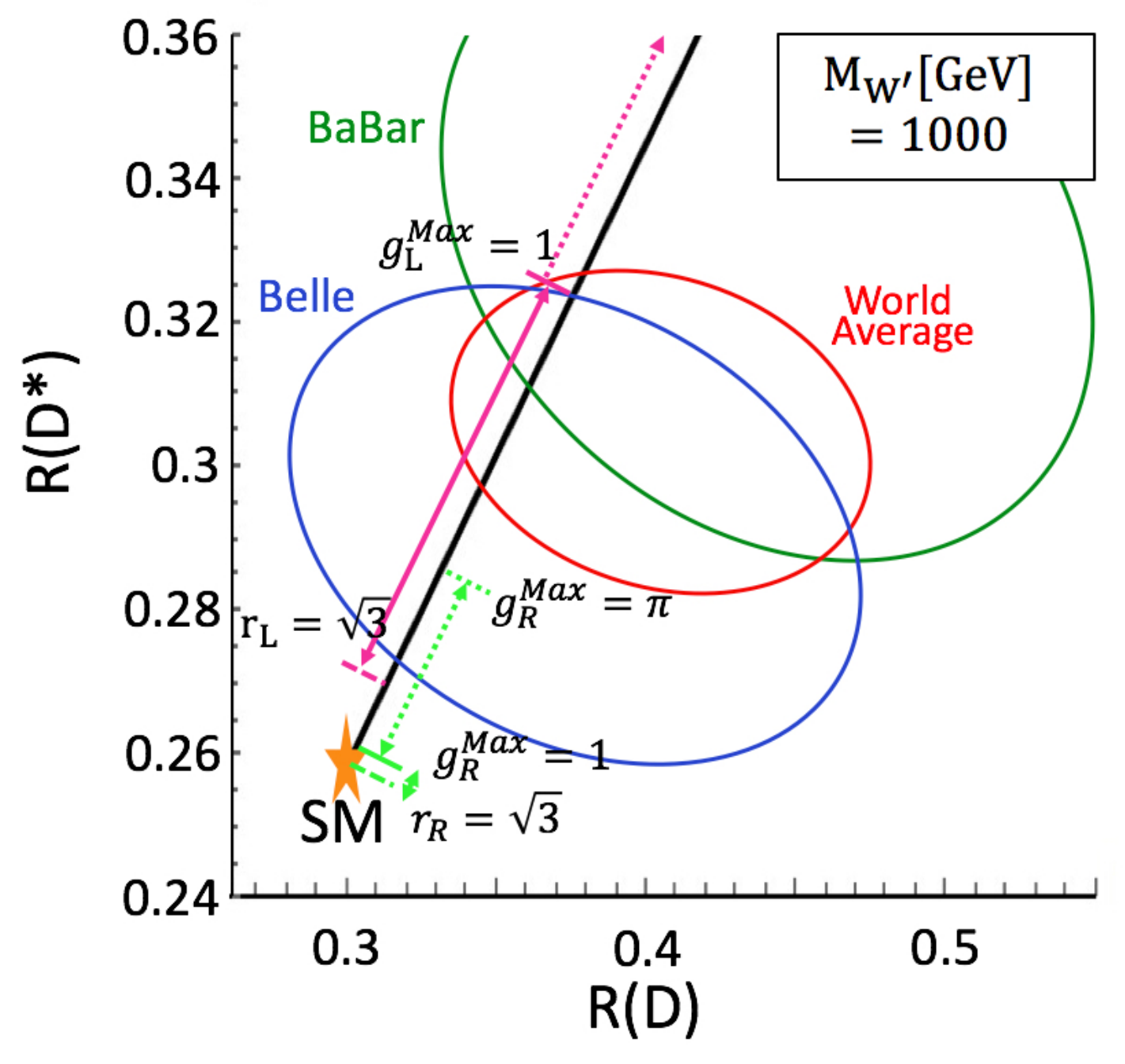}
    \caption{$R(D)$ vs. $R(D^*)$ with the constraint from the charged heavy resonance search at the LHC. 
    $R(D)$ and $R(D^*)$ plane are drawn fixing $W^\prime_I$ masses at 400 GeV, 500 GeV, 750 GeV and 1000 GeV.
The prediction of the $W^\prime_I$ model is on the black line.
The 1 $\sigma$ regions reported by the Belle, BaBar and the HFLAV collaborations are depicted by the same color as in Fig. \ref{Figure-H}. 
The SM prediction for $R(D)$ and $R(D^*)$ is marked with the asterisk. 
On the magenta thick arrows, the $W^\prime_L$ scenario is excluded depending on $g_{L}$ and $g_{L \tau }$, assuming $|g_{L}| \leq 1$. 
The points on the magenta dashed arrows are excluded as far as $|g_{L}| \leq 1$ is required but can be survived allowing $|g_{L}| \leq \pi$. 
The green arrows show the corresponding cases for the $W^\prime_R$.}
    \label{Figure-W}
  \end{center}
\end{figure}

In Fig. \ref{Figure-W}, our predictions on the $R(D)$ and $R(D^*)$ plane are drawn for fixing $W^\prime_I$ masses at 400 GeV, 500 GeV, 750 GeV and 1000 GeV.
The prediction of the $W^\prime_I$ model ($I=L,R$) are along the black line.
The 1 $\sigma$ regions reported by the Belle, BaBar and the HFLAV collaborations are depicted 
by the same color as in Fig. \ref{Figure-H}. 
The SM prediction for $R(D)$ and $R(D^*)$ is marked with the asterisk.

\begin{figure}[h]
  \begin{center}
    \includegraphics[width=7cm]{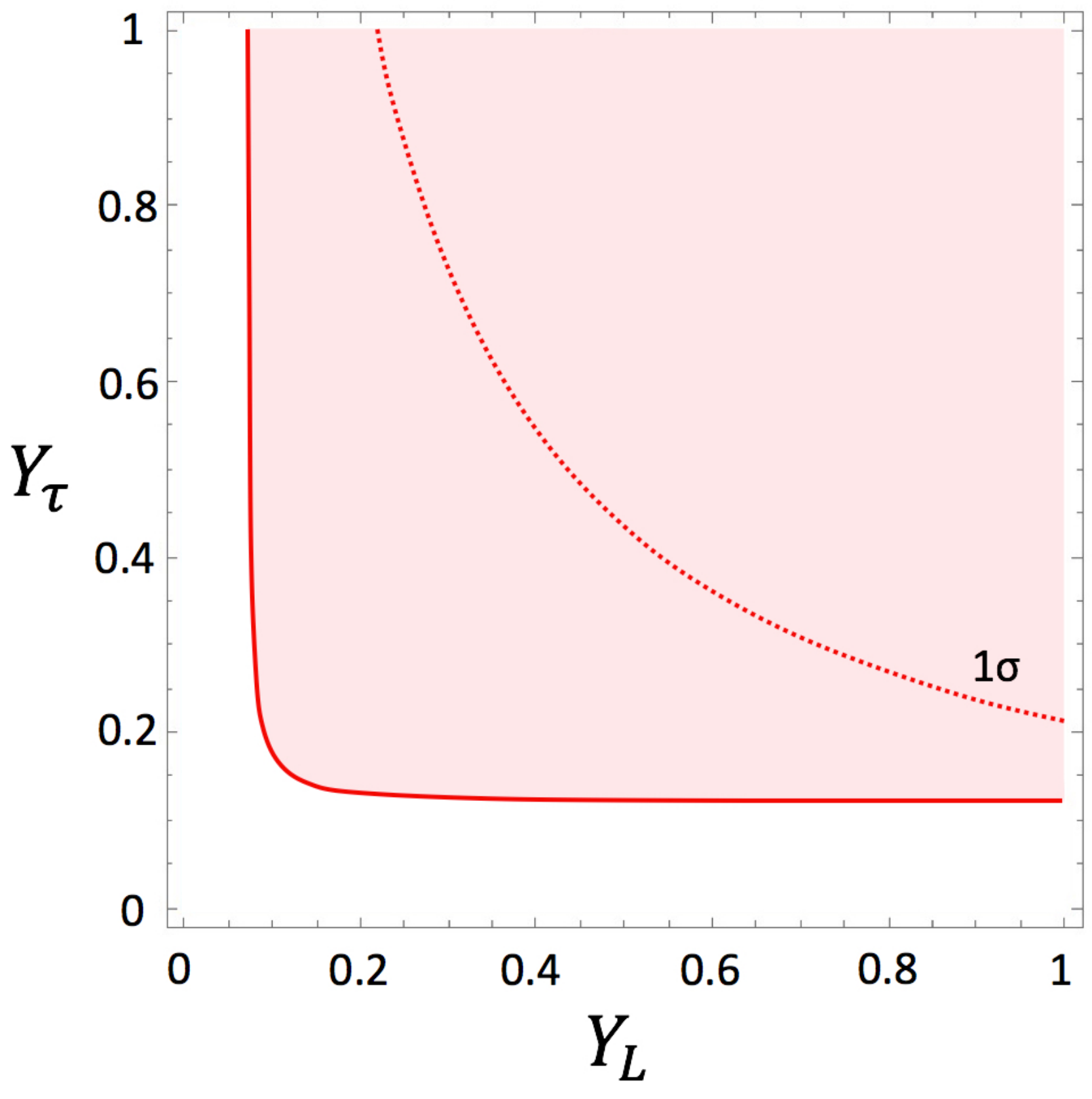}
    \includegraphics[width=7cm]{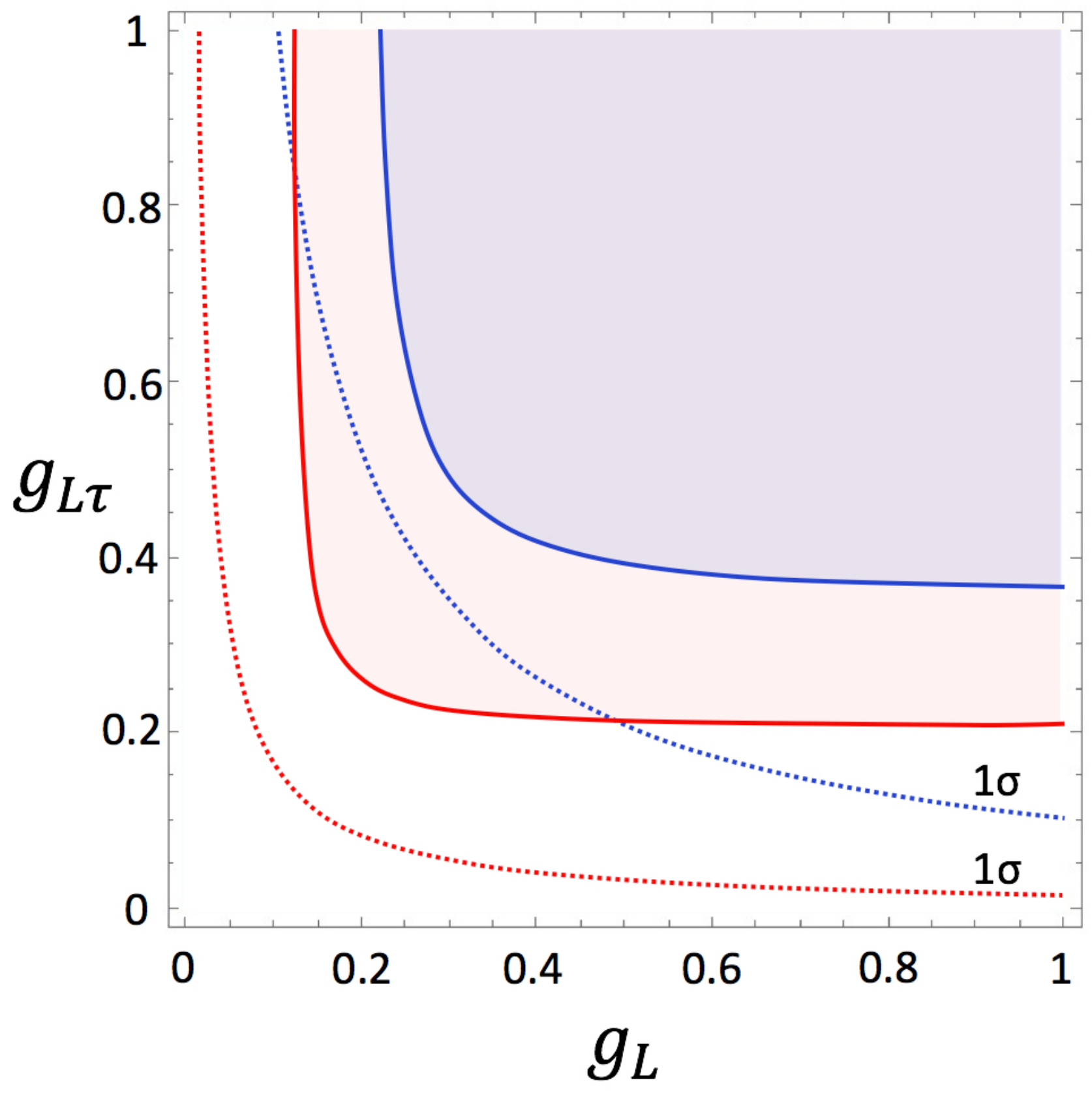}
    \includegraphics[width=7cm]{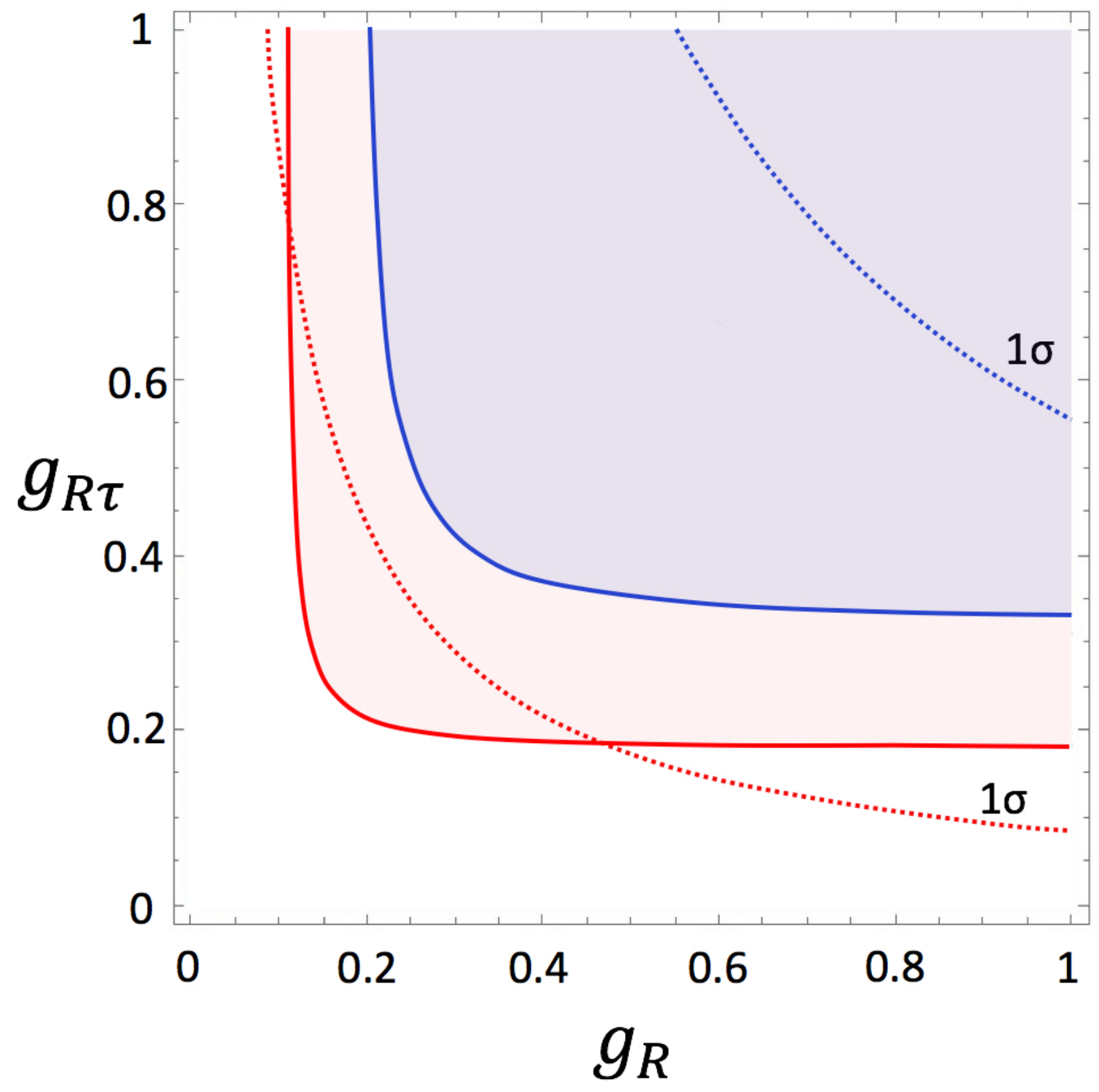}
    \caption{In those figures, the blue (red) lines are for the mass of 1TeV (400GeV).
    The figure on the left (right) above is for the $H_-$($W^\prime_L$) scenario. The last one on the below is for the $W^\prime_R$ scenario.
    The dotted lines express the combination of couplings that accomodate the 1 $\sigma$ of $R(D^{(*)})$. The shaded regions are excluded by the collider bound. 
        }
    \label{fig:coupling} 
  \end{center}
\end{figure}

We show the possibly surviving region taking the allowed range of the couplings $|g_I| \le 1$ into account,
in the magenta arrows ($I=L$) and in the green arrows ($I=R$).
On the magenta thick line, the $W^\prime_L$ scenario is surviving by taking $g_{L}$ and The parameter region on the magenta dashed line are completely excluded when we require $|g_{L}| \leq 1$, while it can be survived 
when we allow $|g_{L}| \leq \pi$.  The green arrows shows the corresponding region for $W_R^\prime$.
 In Fig. \ref{fig:coupling}, we also study the excluded parameter region on the $g_{L(R)}$ and $g_{L(R)\tau}$ plane based on Fig.~\ref{fig:LHCconstraints}. 
  In those figures, the blue(red) lines are for the mass of 1TeV (400GeV).
    The figure on the upper right (bellow) is for the $W^\prime_L$ ($W^\prime_R$) scenario.
    The dotted lines express the combination of couplings that accomodate the 1 $\sigma$ of $R(D^{(*)})$.
     The shaded regions are excluded by the collider bound.
     The parameter region for the $W^\prime_L$ is not constrained at all.
     The parameter region for the $W^\prime_R$ is partially constrained depending on the combination of the couplings.


\section{Summary}
\label{summary}

The discrepancies of $R(D)$ and $R(D^*)$ may be the evidence of new physics behind the SM.
The excesses suggest that there are extra fields that couple to quarks and leptons flavor-dependently and the size of the coupling is not so small compared to the $W$ boson coupling in the SM.
Motivated by this issue, many new physics interpretations have been proposed, and we find that
some of good candidates for the extra fields are charged scalar and charged vector fields. 
Those extra charged fields are predicted by many new physics models beyond the SM.
The charged scalar is, for instance, predicted by many extended SMs with extra SU(2)$_L$-doublet scalar fields.
Extending the SM gauge group is prophetic of some extra massive gauge bosons.
The charged vector can be originated from some extra non-abelian gauge symmetry.

Those good candidates have been studied in flavor physics and collider physics.
The simple setups, where the heavy charged scalars couple to the light quarks and leptons,
have been already excluded, so that some specific structures of the extended SMs are 
required to achieve the explanations. Such a specific setup makes 
it difficult to test indirectly and directly by the experiments.

In this paper, we investigate one observable that does not depend on the detail of the setup, that is, 
the $\tau \nu$ resonance originated from the charged particle. 
We simply consider each minimal setup in the scalar case and in the vector case, 
and discuss the consistency between the explanation of $R(D^{(*)})$ and the latest experimental result at the LHC.
Interestingly, we found most of the parameter region required by the excesses has been already covered by 
the resonance search. If the excesses come from the new interactions, 
the heavy resonances would be relatively light as shown in Fig.~\ref{Figure-H} and Fig.~\ref{Figure-W}.

Our analysis assumes that the heavy resonances only decay into $bc$ and/or $\tau \nu$.
If they decay into the other SM fermions or other particles, the bounds obtained in this paper can be relaxed.
Such a case, however, faces stronger bounds from flavor physics, since the couplings with light quarks and leptons are large.
Therefore, our results would be applicable as long as the decays to $bc$ and/or $\tau \nu$ are dominant.

It may be interesting to study the bound from the $bc$ resonance search. 
In the final state, one or two $b$ quarks appear in association with one $c$ quark.
Currently, the bound from the di-jet resonance is weaker than that of the $\tau\nu$ resonance 
search by 1 or 2 order of magnitude~\cite{CMS:2017xrr,Aaboud:2018fzt}.
If we can identify the $b/c$-flavor jet efficiently, we can set stronger exclusion lines~\cite{CMS:2016knj,ATLctag}.

Finally, we comment on the polarization of $D^*$ in the $B \to D^* \tau \, \nu$ process.
Recently, the Belle collaboration has reported the result on the polarization with high accuracy~\cite{Adamczyk}.
This measurement may be useful to test the new physics scenarios motivated
by the $R(D^{(*)})$ anomaly. In fact, the possible observables 
concerned with this problem are recently discussed~\cite{Bardhan:2016uhr,Azatov:2018knx,Huang:2018nnq}. 
The more general analysis for the $D^*$ polarization and $R(D^{(*)})$,
based on the latest experimental result, will be given near future~\cite{IKOWY}.


\section*{Acknowledgments}


The work of Y. O. is supported by Grant-in-Aid for Scientific research from the Ministry of Education, Science, Sports, and Culture (MEXT), Japan, No. 17H05404.
MT is supported in part by the JSPS Grant-in-Aid for Scientific Research
Numbers~16H03991, 16H02176, 17H05399, 18K03611, and the World Premier
International Research Center Initiative, MEXT, Japan.
The authors thank to Kazuhiro Tobe, Tomomi Kawaguchi, Makoto Tomoto and Yasuyuki Horii  for valuable discussions. 
The authors also thank to David Shih, Matthew Buckley and Poyta Asadi for valuable comments.
\appendix


\end{document}